\documentclass{article}

\usepackage{amsmath}
\usepackage{graphicx}
\usepackage{amsmath}
\usepackage{mathtools}
\usepackage{amsthm}
\usepackage{epstopdf}
\usepackage{pstricks}    %for embedding pspicture.
\usepackage{amssymb}
\usepackage[toc,page]{appendix}
\usepackage{setspace}
\usepackage{tocvsec2}
\usepackage{sectsty}
\usepackage{multicol}% http://ctan.org/pkg/multicol
\usepackage{siunitx}% http://ctan.org/pkg/siunitx
\usepackage{etoolbox} 

\usepackage{titlesec}

\titlelabel{\thetitle.\quad}

\usepackage[labelsep=period]{caption}

\newcommand{\tocless}[2]{\bgroup\let\addcontentsline=\nocontentsline#1{#2}\egroup}

\makeatletter
\newif\if@gather@prefix 
\preto\place@tag@gather{% 
  \if@gather@prefix\iftagsleft@ 
    \kern-\gdisplaywidth@ 
    \rlap{\gather@prefix}% 
    \kern\gdisplaywidth@ 
  \fi\fi 
} 
\appto\place@tag@gather{% 
  \if@gather@prefix\iftagsleft@\else 
    \kern-\displaywidth 
    \rlap{\gather@prefix}% 
    \kern\displaywidth 
  \fi\fi 
  \global\@gather@prefixfalse 
} 
\preto\place@tag{% 
  \if@gather@prefix\iftagsleft@ 
    \kern-\gdisplaywidth@ 
    \rlap{\gather@prefix}% 
    \kern\displaywidth@ 
  \fi\fi 
} 
\appto\place@tag{% 
  \if@gather@prefix\iftagsleft@\else 
    \kern-\displaywidth 
    \rlap{\gather@prefix}% 
    \kern\displaywidth 
  \fi\fi 
  \global\@gather@prefixfalse 
} 
\def\math@cr@@@align{%
  \ifst@rred\nonumber\fi
  \if@eqnsw \global\tag@true \fi
  \global\advance\row@\@ne
  \add@amps\maxfields@
  \omit
  \kern-\alignsep@
  \if@gather@prefix\tag@true\fi
  \iftag@
    \setboxz@h{\@lign\strut@{\make@display@tag}}%
    \place@tag
  \fi
  \ifst@rred\else\global\@eqnswtrue\fi
  \global\lineht@\z@
  \cr
}
\newcommand*{\beforetext}[1]{% 
  \ifmeasuring@\else
  \gdef\gather@prefix{#1}% 
  \global\@gather@prefixtrue 
  \fi
} 
\makeatother

\usepackage[margin=1.2in]{geometry}

\begin{document}

%%%%%%%%%%%%%%%%%%%%%%%%%%%%%%%%%%%%%%%%%%%%%%%%%%%%
%%%%%%%%%%%                                   VARIABLE BOTTOM                                             %%%%%%%%%%%
%%%%%%%%%%%%%%%%%%%%%%%%%%%%%%%%%%%%%%%%%%%%%%%%%%%%

\begin{flushleft}

{\LARGE\textbf{Surface waves over currents and uneven bottom  \\}} \vspace {10mm}
\vspace{1mm} \noindent

{\large \bf Alan C. Compelli$^{a,b,c}$},
{\large \bf Rossen I. Ivanov$^{a,b,*}$},
{\large \bf Calin I. Martin$^{b,c}$},
{\large \bf Michail D. Todorov$^d$} 

\vskip 1cm
\noindent
$^a${\it School of Mathematical Sciences, Dublin Institute of Technology, }{\it Dublin 2, Ireland} 
\hskip 1cm
\\
$^{b}${\it Erwin Schr\"odinger International Institute for Mathematics and Physics, University of Vienna, Vienna, Austria} \\
$^c${\it School of Mathematical Sciences, University College Cork,  }{\it Cork, Ireland} 
\\
$^d${\it Department  of Differential Equations, Faculty of Applied Mathematics and Informatics, }\\ {\it Technical University of Sofia, 8 Kl. Ohridski Blvd., 1000 Sofia, Bulgaria} 
\end{flushleft}

\vskip 1cm

\begin{abstract}
\noindent The propagation of surface water waves interacting with a current and an uneven bottom is studied. Such a situation is typical for ocean waves where the winds generate currents in the top layer of the ocean. The role of the bottom topography is taken into account since it also influences the local wave and current patterns. Specific scaling of the variables is selected which leads to approximations of Boussinesq and KdV types. The arising KdV equation with variable coefficients, dependent on the bottom topography,  is studied numerically when the initial condition is in the form of the one soliton solution for the initial depth. Emergence of new solitons is observed as a result of the wave interaction with the uneven bottom.
\end{abstract}

*Corresponding author. 

E-mail address: rossen.ivanov@dit.ie (R. I. Ivanov)

\section{Introduction}

\hspace{\parindent}Surface waves are examples of geophysical waves found ubiquitously in the oceans. They usually result from distant winds and may travel thousands of miles before they decay or break. 
Typically surface wave heights are 1-2 meters but extreme wave heights approaching nearly 20 meters (excluding rogue waves) have been observed. 

The complete dynamics of the fluid, in many situations, can be recovered from the surface wave motion. Therefore the wave propagation is an important research topic, with possible coastal 
and oceanic engineering applications in areas like oil platform stability and designs of structures like  breakwaters and artificial reefs.  

Various factors play a role in the formation and the dynamics of the surface waves. Some of these are rotational effects (for example currents) and the bottom topography. 
The propagation of waves over a current field is, in general, a complicated phenomenon and an active research topic.  
Currents exist due to many factors but primarily due to wind and tides. One of the most significant currents is the Equatorial Undercurrent (EUC) which can be observed at approximately 
20 degrees south of the Equator. The EUC flows eastwards, at a speed of around one meter per second, one hundred meters below the ocean surface and is approximately 300 kilometers wide. 
Some more details about the complexity of the dynamics of EUC which pertain to the analysis of solutions can be found in the papers by Constantin and Johnson (2016a, 2017b).
A general overview on mathematical tools employed for the study of problems in physical oceanography is given in Johnson (2018). 
We would also like to point out the recent emergence of
solutions modelling large-scale ocean currents with distinctive persistence patterns like gyres (Constantin and Johnson, 2017a) and the Antarctic Circumpolar Current (Constantin and Johnson, 2016b). An analysis of a solution describing nonlinear surface waves propagating zonally on a zonal current in the presence of Coriolis effects and exhibiting two modes of wave motion was presented by Constantin and Monismith (2017).

For the aim of our investigations we consider two-dimensional 
flows in the $f$-plane approximation. Such an approach is reasonable since the Equator acts as a wave guide (Fedorov and Brown, 2009) and the depth-dependent currents are confined to a shallow 
near-surface layer. For a rigorous existence proof of rotational water flows in the $f$-plane approximation we refer the reader to the paper by Constantin (2013); see also the paper by Martin (2018).

In the review chapters by Peregrine (1976) and Jonsson (1990), the physical circumstances in which interactions among water waves and currents occur are discussed and also mathematical models of 
these interactions are outlined. Thomas (1981, 1990) studied the horizontal velocity distribution and surface elevation in a wave-current environment. A method has been developed for the
measurement of the strength of the wave-current interaction by Thomas and Klopman (1997). A variety of other aspects has been an active research topic 
(Teles da Silva and Peregrine, 1988; Constantin and Strauss, 2004; Constantin and Escher,  2004 and 2011; Constantin et al., 2006; Henry, 2013; Constantin and Johnson, 2015 and 2017a,b;  
Escher et al., 2016). Recent developments in the mathematical aspects of wave-current interactions are presented in the monograph by Constantin (2011). 
 
In physical reality, the seabed can be a complex undulating structure with trenches, underwater mountains and sediment, not to mention the multitude of marine flora and fauna. 
In some parts of the ocean, due to volcanic activity, the seabed is not even a stationary structure. Numerous studies are dedicated to the wave motion over an uneven bottom 
(Johnson, 1973a,b, 1997; Nachbin, 2003; Craig et al.,2005a; Dejak and Sigal, 2006; de Bouard et al., 2008; Compelli et al., 2017; Nachbin and Ribeiro-Jr.,  2018). 
In the majority of the existing publications however, the fluid is irrotational, i.e. the underlying current is not taken into account.

In our studies we are going to apply the Hamiltonian approach to derive models for surface wave propagation in the presence of a current and of an uneven bottom. The significance of the work of Hamilton in developing Hamiltonian mechanics in the early 19th century became apparent in the context of fluid systems following the publication of Vladimir Zakharov (1968) with additional constructions by Milder (1977) and Miles (1977). In his work Zakharov demonstrated the Hamiltonian structure of an infinitely deep vorticity-free single fluid system. Further significant studies which followed on include the works of Benjamin and Olver (1982), Craig and Groves (1994), Craig et al. (2005b) among others.
The Hamiltonian approach in the analysis of irrotational single media systems, was then extended to include systems with nonzero vorticity. For instance a rotational single fluid system, of finite-depth, was shown by Constantin et al. (2008) to exhibit \textit{nearly} canonical Hamiltonian structure and Wahl{\'e}n (2007) 
showed that a fully canonical structure could indeed be achieved via a variable transformation. Moreover, Constantin et al. (2016) showed that wave-current interactions in stratified rotational flows also possess a Hamiltonian formulation.

Our aim is to derive and investigate a model, which takes into account both the effect of the current and the non-flat bottom. Such a situation is common for ocean waves and waves over the EUC are just one particular realistic example. To fulfil our goal we show first that the governing equations for gravity water flows in the $f$-plane approximation (near the Equator) that exhibit constant (non-vanishing) vorticity and propagate over a non-flat bed can be written in Hamiltonian
form. Subsequently, we derive from the Hamiltonian formulation a small amplitude and a long-wave approximation, typically related to the Boussinesq and KdV propagation regimes. 
We use several mathematical results explained previously in Constantin et al. (2016) and Compelli et al. (2018). We would like to mention that the Hamiltonian formulation for two-layered
stratified water flows over a flat bed and displaying piecewise constant vorticity in the equatorial $f$-plane approximation was accomplished by Ionescu-Kruse and Martin (2017).

\section{System set-up}
\begin{figure}[h!]
\begin{center}
\includegraphics[totalheight=0.43\textheight]{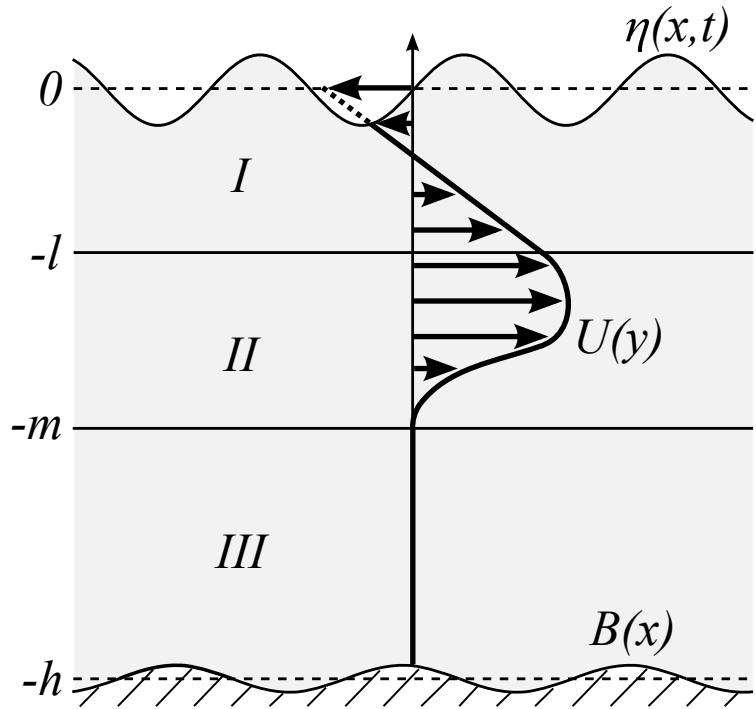}
\caption{Set-up for the variable-bottom system}
\label{fig:figure_system}
\end{center}
\end{figure}

\hspace{\parindent}We consider an incompressible fluid system consisting of a single inviscid medium as shown in Fig. \ref{fig:figure_system}. The velocity profile of the underlying current is given in the same figure. 

The average surface level is taken at $y=0$, and the average bottom level at $y=-h$.  The fluid domain  $\Omega$ is bounded from above by the surface wave $y= \eta(x,t)$ and 
underneath by a stationary impermeable bottom described by the function $y=B(x)$. Since the average bottom level is $y=-h,$ it is suitable to introduce also the bottom elevation function $\beta(x)$ via
\begin{equation}
\label{TheLocalDepth}
B(x)=:-h+\beta(x).
\end{equation}

Given the average values for the surface elevation and the bottom level, we have 
\begin{equation}
\label{IntEta}
\int\limits_{\mathbb{R}} \eta(x,t)dx =0, \qquad \int\limits_{\mathbb{R}} \beta(x,t)dx =0.
\end{equation}

The typical example for a physical system is the wave motion along the Equator over the Equatorial Undercurrent. On the surface the current velocity is negative (westward), 
as the depth increases it becomes positive (eastward), while below a certain depth there is no current. Given constants $\gamma$ and $\kappa$,
a realistic model of such a current is 
\begin{equation}
\label{CurrentProfile}
U(y)=
\begin{dcases}     
        \gamma y+\kappa\mbox{ for }\eta\ge y \ge -l\mbox{ (layer I)}
        \\
        U_{II}(y) \mbox{ for }-l\ge y \ge -m\mbox{ (layer II)}
        \\
        0\mbox{ for }-m\ge y \ge B(x)\mbox{ (layer III)}
\end{dcases}
\end{equation}
where $l$ and $m$ are some positive constants describing the depths at which significant changes in the current profile occur, cf. Fig. \ref{fig:figure_system}. 
The current \eqref{CurrentProfile} is subjected to the constraints
\begin{align}
&U_{II}(-l)=-\gamma l+\kappa\\
\beforetext{and}&U_{II}(-m)=0.
\end{align}

Note that $U(0)=\kappa$ is the current velocity on the surface. Typical values are $\kappa=-2$ m s$^{-1}$. Since the wind usually moves the top 20 m of the sea surface, assuming that $U(-20)\approx 0$ we can make the estimate $\gamma \approx  -10^{-1}$ rad s$^{-1}$.

The subscript notation $s$ will be used to refer to evaluation on the surface $y=\eta(x,t)$, while $b$ will mean evaluation on the bottom $y=B(x)$.

\section{Governing equations}
\hspace{\parindent}Denoting the velocity field using $(u,v)$,
the motion of the water flows we consider is governed by the equation of mass conservation
\begin{equation}\label{masscons}
{ u}_x+ {v}_y=0,\,\,
\end{equation}
and Euler's equations in the presence of the Coriolis force 
\begin{equation}
 \left\{\begin{array}{lcl}
        { u}_t+{u}{u}_x+{v}{u}_y+2\omega {v} &=& -{\frac{1}{\rho}}P_x,\\
         {v}_t+{u}{v}_x+{v}{v}_y -2\omega {u} &=& -{\frac{1}{\rho}}P_y-g,
        \end{array}\right.
\end{equation}
where $P=P(x,y,t)$ denotes the pressure, $\omega$ is the rotational speed of Earth, $g$ is the gravitational acceleration and ${\rho}$ denotes the density of the fluid and is assumed to be constant.

The vorticity of the water flow, defined as the curl of the velocity field $(u,v)$, takes in our case the shape
\begin{equation}\label{vorticity}
{u}_y-{v}_x =U'(y).
\end{equation}
Relations \eqref{CurrentProfile} and \eqref{vorticity} imply that the vorticity equals the constant $\gamma$ in the top layer, while vanishing in the bottom layer.
\begin{subequations}\label{stream_and_potential}
From the equation of mass conservation \eqref{masscons} we deduce the existence of a stream function, denoted $\psi(x,y,t)$, which satisfies
\begin{align}
u(x,y,t)= \psi_y(x,y,t)\quad {\rm and}\quad  v(x,y,t)= -\psi_x(x,y,t)
\end{align} and is defined up to an addition of an arbitrary function of $t$.
Moreover, relation \eqref{vorticity} entails the existence of a (generalized) velocity potential $\varphi(x,y,t)$ satisfying
\begin{align} \label{pot}
u(x,y,t)=\varphi_x(x,y,t) +U(y)\quad {\rm and}\quad  v(x,y,t)= \varphi_y(x,y,t).
\end{align}
\end{subequations}

As a consequence of the equation of mass conservation \eqref{masscons} and the equations from \eqref{pot} we obtain
\begin{equation}\label{Delta_phi}
\Delta \varphi(x,y,t)=0,
\end{equation}
for all $(x,y)$ that lie in the fluid domain and for all $t$.

For simplicity we assume that, for any $y$ and $t$, the functions $x\rightarrow\eta(x, t)$ and $x\rightarrow\varphi(x, y, t)$ belong to the Schwartz class $\mathcal{S}(\mathbb{R})$ of smooth, 
rapidly decreasing functions. 
The assumption, of course, implies that, for large absolute values of $x$, the wave attenuates, and  disappears at infinity:
\begin{align}
\label{WaterWaveAssumps1}
&\lim_{|x|\rightarrow \infty}\eta(x,t)=0\\
\label{WaterWaveAssumps2}
\beforetext{and}&\lim_{|x|\rightarrow \infty}\varphi(x,y,t)=0,
\end{align}
the same properties being true for the $x$ derivatives of $\eta$ and $\varphi$.

The equations of motion are completed by the boundary conditions of the free surface and on the bottom boundary.
We start with the kinematic boundary conditions which state the impermeability of each of the two boundaries: once a particle is on one of the boundaries, it remains there at all times $t$.
\begin{subequations}
The kinematic boundary condition on the surface is given as
\begin{align}
\label{KBC_VarBottSystem}
\eta_t(x,t)=v(x,\eta(x,t),t)-u(x,\eta(x,t),t) \eta_x (x,t),
\end{align}
for all $x$ and $t$.
In the same vein, the kinematic boundary condition on the bed $y=B(x)$ is
\begin{align}
 v(x,B(x),t)=u(x,B(x),t)B^{\prime}(x),
\end{align}
for all $x$ and $t$.
\end{subequations}

The kinematic boundary condition on the free surface \eqref{KBC_VarBottSystem} can be expressed in terms of the stream function as (Johnson, 1997; Constantin et al. 2008)
\begin{equation}
\label{KBC_VarBottSystem_alt}
\eta_t=-(\psi_{x})_s-(\psi_{y})_s\eta_x
\end{equation}
and in terms of the velocity potential as
\begin{equation}
\label{KBC_VarBottSystem_alt2}
\eta_t=(\varphi_y)_s-\left((\varphi_x)_s+\gamma \eta+\kappa\right)\eta_x.
\end{equation}

Furthermore, on the surface we impose the dynamic boundary condition that decouples the motion of the water from the motion of the air above by assuming
the pressure equals the constant atmospheric pressure, $P_{atm}$, that is
\begin{equation}\label{atm}
 P=P_{atm}\,\,{\textrm on}\,\, y=\eta (x,t).
\end{equation}

Euler's equations can be recast by means of the stream function and the generalized velocity potential as
\begin{equation}
\nabla\left[{\varphi}_t+\frac{1}{2}|\nabla{\psi}|^2+\frac{P}{\rho}-({\gamma}+2\omega){\psi}+g y\right]=0.
\end{equation}
We have therefore
$$\varphi_{t}+\frac{1}{2}|\nabla\psi|^2 -(\gamma+2\omega)\psi +\frac{P}{\rho}+gy=f(t)\quad{\rm in \,\, the \,\, fluid,}$$
where $f(t)$ is an arbitrary function, reflecting the gauge freedom in the choice of the potential. Changing $\varphi$, if necessary, by a time-dependent factor and making use of \eqref{atm} we obtain
\begin{equation}\label{Btop}
 \varphi_{t}+\frac{1}{2}|\nabla\psi|^2-(\gamma+2\omega)\psi+g y=0\quad{\rm on}\quad y=\eta(x,t).
\end{equation}

In other words, we can write the so-called Bernoulli condition on the surface as

\begin{equation} \label{varphi_t}
(\varphi_{t})_s +\frac{1}{2}|\nabla \psi|_s^2-(\gamma+2\omega)\chi+g\eta=0
\end{equation}
where $\chi(x,t):=\psi(x,\eta(x,t),t)$ is the stream function on the surface.

Equations \eqref{KBC_VarBottSystem_alt2} and \eqref{varphi_t} provide the time evolution of the surface variables $\eta(x,t)$ and $(\varphi)_s \equiv \varphi(x, \eta(x,t), t)$ which, as we shall see in the next section, play the key roles in the Hamiltonian formulation of the problem.  

\section{Hamiltonian formulation}

\hspace{\parindent}The total energy written as the sum of kinetic and potential energy contributions is
\begin{equation}
H =\frac{1}{2}\rho\int\limits_{\mathbb{R}} \int\limits_{B}^{\eta}  (u^2+v^2)dy dx\\+\rho g\int\limits_{\mathbb{R}} \int\limits_{B}^{\eta}  y \, dy dx. \end{equation}
This energy functional written in terms of the appropriate variables will serve as a Hamiltonian for the system under consideration. The energy of the current in the bottom layer is zero, 
in the middle layer ($-m\le y\le -l$) is constant, so the only physical contribution to $H$ is coming from the top layer.  
With computations, similar to those in Constantin et al. (2016), Compelli (2016), Compelli and Ivanov (2017) and Compelli et al. (2018), one can represent the energy functional in the form 
\begin{equation} \label{H}
H[\xi, \eta] = H_0[\xi, \eta]
-\int\limits_{\mathbb{R}}   (\gamma \eta+\kappa)\xi\eta_x dx 
+\frac{1}{6}\rho\gamma^2\int\limits_{\mathbb{R}} \eta^3  dx+
\frac{1}{2}\rho\gamma\kappa\int\limits_{\mathbb{R}}\eta^2 \, dx
\end{equation} where 
\begin{equation}
\begin{split}
H_0 [\xi, \eta]=  &\frac{1}{2}\rho\int\limits_{\mathbb{R}}  \int\limits_{B}^{\eta} |\nabla {\varphi}|^2 dy dx +\frac{1}{2}\rho g \int\limits_{\mathbb{R}}\eta^2 dx= \frac{1}{2\rho}\int\limits_{\mathbb{R}}  \xi  G(\beta,\eta) \xi \, dx +\frac{1}{2}\rho g \int\limits_{\mathbb{R}}\eta^2 dx
\end{split}
\end{equation} is the energy of a system without a current (Compelli et al. (2018)). The other notations are as follows: 
\begin{equation}
\xi(x,t):=\rho \varphi(x,\eta(x,t),t), \qquad   \phi(x,t):=\varphi(x,\eta(x,t),t), 
\end{equation}
and the Dirichlet-Neumann operator $G(\beta,\eta)$  for equation (\ref{Delta_phi}) for the fluid domain under consideration is given by
\begin{equation}
 G(\beta,\eta)\phi =-\eta_x (\varphi_x)_s+(\varphi_y)_s =(-\eta_x, 1) \cdot (\nabla \varphi)_s=\sqrt{1+\eta_x^2}\,\,  {\bf n}_s \cdot (\nabla \varphi)_s
\end{equation}
where  ${\bf n}_s = (-\eta_x, 1)/\sqrt{1+\eta_x^2}$ is the outward-pointing unit normal vector to the water surface (Craig et al., 2005a).

Using the relations (Constantin et al. 2008; Constantin et al. 2016; Compelli et al. 2018)
\begin{equation}
\frac{\delta H_0}{\delta \xi}= (\varphi_y)_s-(\varphi_x)_s\eta_x, \qquad \frac{\delta H_0}{\delta \eta}= - \rho \big((\varphi_y)_s-(\varphi_x)_s \eta_x\big)(\varphi_y)_s    +\frac{1}{2} \rho |\nabla \varphi|_s^2  +\rho g \eta
\end{equation}
we can write equation (\ref{KBC_VarBottSystem_alt2}) in the form 
\begin{equation}\label{H1}
\eta_t=\frac{\delta H}{\delta \xi}
\end{equation}
and equation (\ref{varphi_t}), noting that $\xi_t=  \rho(\varphi_t)_s   +\rho (\varphi_y)_s \eta_t,$  $\xi_x=  \rho (\varphi_x)_s   +\rho (\varphi_y)_s \eta_x,$ in the form

\begin{equation}
\label{H2}
\xi_t=-\frac{\delta H}{\delta\eta}+\rho(\gamma+2\omega)\chi,
\end{equation}
where the Hamiltonian $H$ is given by equation (\ref{H}). The system \eqref{H1}-\eqref{H2} can be written in a canonical Hamiltonian form in terms of the variables $\eta$ and 
$$\zeta= \xi+\frac{\rho(\gamma+2\omega)}{2}\int_{-\infty}^{x}\eta(x^{\prime},t)\,dx^{\prime}$$ (Wahl{\'e}n, 2007); however, 
for our purposes we prefer the following form. Noting that $\chi_x=(\psi_x)_s+ (\psi_y)_s\eta_x=-\eta_t$, as follows from equation (\ref{KBC_VarBottSystem_alt}), 
and introducing the new variable 
\begin{equation}
\mathfrak{u}=\xi_x\notag,
\end{equation}
we can transform the system  \eqref{H1}-\eqref{H2} into

\begin{equation}
\label{HF}
%\begin{dcases}
    \mathfrak{u}_t= -\left(\frac{\delta H}{\delta \eta}\right)_x-\rho(\gamma+2\omega)\eta_t, \qquad
    \eta_t= - \left(\frac{\delta H}{\delta \mathfrak{u}}\right)_x.
%\end{dcases}
\end{equation}

Thus, the dynamics now is determined only by the Hamiltonian $H$ written in terms of $\eta$ and $\mathfrak{u}$. In what follows we will determine an expression for $H[\eta, \mathfrak{u}]$ for a specific wave propagation regime.

\section{The Boussinesq and KdV approximations}

\hspace{\parindent}We will make now some further assumptions about the scales of the quantities involved. Introducing the usual small scale parameters 
\begin{equation}
\varepsilon=\frac{a}{h}\quad\mbox{and}\quad\delta=\frac{h}{\lambda},\notag
\end{equation}
where $a$ is the typical wave amplitude and $\lambda$ is the typical wavelength, we consider the long-wave and shallow water scaling regime. Then clearly $\eta/h$ is of order $\varepsilon$. The operator  $h \partial$ (where $\partial := \partial/\partial x$) has an eigenvalue of order $ 2\pi h/\lambda,$ that is of order $\delta. $ The quantity $\mathfrak{u}=\xi_x$ has the meaning of a velocity multiplied by $\rho$, and therefore is of order $\varepsilon$; see more details in Compelli et al. (2018).

Furthermore, the bottom variations are considered small, such that $X=\varepsilon x$ is of order 1. It is assumed that $\beta(x)$ can be of the order of $h$ provided that the bed stays away from the surface, that is $|\beta_{\text{max}}|/h$ is of $\mathcal{O}(1)$ with $|\beta(x)|/h < 1.$ This way one can expand in powers of $\beta/h .$ We introduce the local depth $$b(X):= h-\beta(\varepsilon x)>0.$$  

$b(X)$ is of order 1. Most importantly, $\varepsilon \sim \delta^2$ which usually leads to the Boussinesq and KdV propagation regimes.
The Dirichlet-Neumann operator $G(\beta, \eta)$ expanded to the given order ($\varepsilon \sim \delta^2$), is evaluated in Compelli et al. (2018). 
The Hamiltonian \eqref{H} becomes (rescaled by $\varepsilon^2$ as an overall scale factor) 
\begin{multline}
H[\eta, \mathfrak{u}]=\frac{1}{2\rho}\int_{\mathbb{R}}  \mathfrak{u}\left(b(X)  +\delta^2\frac{1}{3}\partial b^3 (X) \partial  + \varepsilon \eta \right) \mathfrak{u} dx
+\frac{1}{2}\rho (g+\gamma\kappa)\int_{\mathbb{R}} \eta^2 dx +\kappa\int\limits_{\mathbb{R}} \eta \mathfrak{u} dx  \\
+\frac{1}{2}\varepsilon\gamma\int\limits_{\mathbb{R}} \eta^2 \mathfrak{u} dx +\frac{1}{6}\varepsilon\rho\gamma^2\int\limits_{\mathbb{R}} \eta^3  dx
+ \mathcal{O}(\varepsilon^2).
\end{multline}

The Boussinesq-type equations of motion (Boussinesq, 1872) are determined by \eqref{HF}:
\begin{equation}
\rho(\eta_t+\kappa\eta_x)=-\left(b \mathfrak{u}  \right)_x
-\delta^2\frac{1}{3} b^3 \mathfrak{u}_{xxx}  
-\varepsilon\left( \mathfrak{u} \eta \right)_x
- \varepsilon\rho\gamma\eta\eta_x
\end{equation}
and
\begin{equation}
\mathfrak{u}_t+\kappa\mathfrak{u}_x+\Gamma\eta_t=-\rho (g+\gamma\kappa) \eta_x-\frac{\varepsilon}{\rho}\mathfrak{u}\mathfrak{u}_x 
-\varepsilon\gamma(\eta\mathfrak{u})_x   
-\varepsilon\rho\gamma^2\eta\eta_x,
\end{equation}
where $\Gamma= \rho(\gamma+ 2 \omega)$.
This system in the irrotational case and with a flat bottom ($b$=const.) is integrable. It is known as the Kaup-Boussinesq system (Kaup, 1975), the details of which can be found in Compelli et al. (2018).

In the leading order
\begin{equation}\label{lo}
\begin{split}
&\rho(\eta_t+\kappa\eta_x)=-\left(b \mathfrak{u}  \right)_x\\
\beforetext{and}&\mathfrak{u}_t+\kappa\mathfrak{u}_x +\Gamma\eta_t=-\rho (g+\gamma\kappa) \eta_x.
\end{split}
\end{equation}

The solutions for $\eta$ and $\mathfrak{u}$ can be deduced from the form
\begin{equation}\label{los}
\begin{split}
        \eta(x,t)&=\eta_0e^{ik(x-c(X)t)}\\
\beforetext{and}        \mathfrak{u}(x,t)&=\mathfrak{u}_0e^{ik(x-c(X)t)}
\end{split}
\end{equation}
where $k=2\pi/\lambda$ is the wave number and $c(X)$ is the wave speed. Note that $c$ depends on the ``slowly varying'' variable $X$. From \eqref{lo}-\eqref{los} it is straightforward to obtain the following quadratic equation for the wavespeed $c$:

\begin{equation}
(c-\kappa)^2 +b(\gamma+ 2 \omega)c -(g+\gamma\kappa)b =0.
\end{equation}
This can be written as
\begin{equation}
(c-\kappa)^2
+(\gamma+2\omega) b  (c-\kappa)-(g-2\omega \kappa )b 
 =0.
\end{equation}

Since the Earth's rotation angular velocity is
$\omega\approx7.3\times10^{-5}$ rad s$^{-1}$ and $\kappa$ is usually of a magnitude of several m s$^{-1}$, then $\omega \kappa \ll g$ giving the quadratic equation for $(c-\kappa)$
\begin{equation}
\label{cminuskappa}
(c-\kappa)^2
+(\gamma+2\omega) b  (c-\kappa)-gb 
 =0.
\end{equation}
The solutions are 

\begin{equation} \label{c}
c(X)=\kappa+\frac{1}{2}\left(-(\gamma+2\omega) b(X)\pm\sqrt{(\gamma+2\omega)^2 b^2(X) +4gb(X) }\right).
\end{equation}

For example, for the EUC when $\gamma=-0.1 $ rad s$^{-1}$, $\kappa=-2$  m s$^{-1}$ we see that $\omega$ is negligible and the values for the left and right running speeds when $b=h=4000$ m  are $c_+\approx 480$ m s$^{-1}$ and $c_-\approx -83$ m s$^{-1}$. For a depth $b=500$ m we have $c_+\approx 97$ m s$^{-1}$ and $c_-\approx -51$ m s$^{-1}$. It is evident also that $c(X)$ is usually much bigger than $\kappa$. Another observation is that the presence of vorticity changes considerably the wave speed, which for the irrotational case ($\gamma=0,$ $\kappa=0$, $b=4000$ m) for example, is  $c_{\pm}\approx \pm 198$ m s$^{-1}$.

As in the irrotational case (Compelli et al., 2018), in addition to the variable $X$, we introduce the characteristic variable in the form
\begin{equation}
\theta=\frac{1}{\varepsilon}R(X) - t,
\end{equation}
where $R(X)$ is some function to be determined. The $(x,t)$ coordinate partial derivatives change according to
\begin{align}
    \partial_x&\equiv R'(X) \partial_{\theta}+\varepsilon\partial_X\notag\\
\beforetext{and}    \partial_t&\equiv-\partial_{\theta}\notag.
\end{align}

The equations then can be transformed from $(x,t)$ variables to the slow variables $(\theta, X).$   The leading order of the obtained equations, like in the irrotational case (Compelli et al., 2018) suggests that the appropriate choice for $R(X)$ requires \begin{equation} \label{r}
R'(X) \equiv\frac{1}{c(X)}.
\end{equation} 
This way, of course, two sets of equations arise (for the left and for the right running waves).

The equations written in terms of the new variables are

\begin{equation}
\rho\left((-c+\kappa)  \eta_{\theta}+\varepsilon c\kappa\eta_X\right)
=-  \left(b\mathfrak{u}  \right)_{\theta}-\varepsilon c\left(b\mathfrak{u}  \right)_X
-\delta^2\frac{1}{3} b^3 \frac{1}{c^2} \mathfrak{u}_{\theta\theta\theta}  
-\varepsilon   \left( \mathfrak{u} \eta \right)_{\theta}
- \varepsilon \rho\gamma\eta\eta_{\theta}+\mathcal{O}(\varepsilon^2)
\end{equation}
and
\begin{multline}
(-c+  \kappa) \mathfrak{u}_{\theta}+\varepsilon c\kappa\mathfrak{u}_X -\rho c (\gamma+2\omega)\eta_\theta\\
=-\rho (g+\gamma\kappa)   \eta_{\theta}-\varepsilon c\rho (g+\gamma\kappa) \eta_X-\varepsilon\frac{1}{\rho } \mathfrak{u}\mathfrak{u}_{\theta} 
-\varepsilon \gamma  (\eta\mathfrak{u})_{\theta}
-\varepsilon \rho\gamma^2  \eta\eta_{\theta}  +\mathcal{O}(\varepsilon^2).
\end{multline}
From the first equation
\begin{equation}
\mathfrak{u}_{\theta}
=\frac{\rho}{b}\left((c-\kappa)  \eta_{\theta}-\varepsilon c\kappa\eta_X\right)
-\varepsilon \frac{c}{b}\left(b\mathfrak{u}  \right)_X
-\delta^2 \frac{b^2}{3c^2} \mathfrak{u}_{\theta\theta\theta}  
-\varepsilon \frac{1}{b}  \left( \mathfrak{u} \eta \right)_{\theta}
- \varepsilon \frac{\rho\gamma}{b}\eta\eta_{\theta}+\mathcal{O}(\varepsilon^2).
\end{equation}
So in the leading order
\begin{equation}
\mathfrak{u}
=\frac{\rho}{b}(c-\kappa)  \eta
\end{equation}
and thus 

\begin{equation}
\mathfrak{u}_{\theta}
=\frac{\rho}{b}\left((c-\kappa)  \eta_{\theta}-\varepsilon c\kappa\eta_X\right)
-\varepsilon \frac{\rho c}{b}\left((c-\kappa)\eta  \right)_X
-\delta^2 \frac{\rho b(c-\kappa)}{3c^2} \eta_{\theta\theta\theta}  
-\varepsilon \frac{\rho(c-\kappa)}{b^2}  2 \eta \eta _{\theta}
- \varepsilon \frac{\rho\gamma}{b}\eta\eta_{\theta}+\mathcal{O}(\varepsilon^2).
\end{equation}

Therefore the second equation can be written with $ \mathfrak{u}$ excluded as

\begin{multline} \label{KdV}
\varepsilon \frac{c^2}{b} \left[2(c-\kappa)+ b(\gamma+2\omega)\right] \eta_X
+
\varepsilon  \left[ \frac{c^2 c_X}{b} -\frac{\kappa c(c-\kappa)b_X}{b^2} \right]\eta
\\
+\delta^2\frac{(c-  \kappa)^2 b}{3 c^2}  \eta_{\theta\theta\theta}  + 
 \varepsilon  \frac{3(c-\kappa)^2+3\gamma b(c-\kappa)+\gamma^2 b^2}{b^2}  \, \eta\eta_{\theta}
=0.
\end{multline}

This is a KdV-type equation (Korteweg and de Vries, 1895) with variable coefficients. The coefficients depend on functions, slowly varying with $X.$  The KdV equation with constant coefficients is integrable, however it seems that equation \eqref{KdV} can not be transformed to any integrable forms of the KdV equation. 

For the irrotational case with $\kappa=0,$ $b=c^2/g$ the equation reduces to

\begin{equation}
\varepsilon (2c  \eta_X+ c_X \eta)
+\delta^2 \frac{c^2}{3g^2  }  \eta_{\theta\theta\theta}  + 
 \varepsilon\frac{3g }{c^2}  \eta\eta_{\theta}
=0.
\end{equation}

This is the equation derived by Johnson (1973a,b; 1997). Therefore equation \eqref{KdV} is a generalization of Johnson's equation.

We observe that $X$ plays the role of the time-like variable in the usual KdV setting and $\theta$ - the space-like variable.  This surprising outcome is due to the fact that the original $(x,t)$ variables are both of order $1/\varepsilon$ (while combinations like the characteristics $x-c(X) t$ are of order 1) and at leading order both $x$ and $t$ can measure time:

\begin{equation} \label{t}
t=\frac{1}{\varepsilon}R(X) - \theta\approx \frac{1}{\varepsilon}\frac{X}{\bar{c}(X)}+\mathcal{O}(1)= \frac{x}{\bar{c}(X)}+\ldots,
\end{equation} 
where $\bar{c}$ is some ``average'' propagation speed .\footnote{From \eqref{r} $\bar{c}$ can be defined on the interval $[0, X]$ as $\bar{c}(X) =\left[\frac{1}{X}\int _{0}^{X}\frac{d \tilde{X}}{c(\tilde{X}) } \right]^{-1} .$  }

\section{Wave propagation}

\hspace{\parindent}In this section we study numerically the wave propagation, as governed by eq. \eqref{KdV}. We assume for simplicity that $\kappa=0$ since it is much smaller than $c$ (indeed, $\kappa$ is usually $1-2$ m s$^{-1}$, while $c$ is usually tens to hundreds m s$^{-1}$).  The proper scale $\varepsilon \sim \delta^2$ means that for depths of magnitude $h=200$ m and wave amplitudes $\eta \sim 20$ m we have wavelengths $\eta/h \sim (h/\lambda)^2$ or $\lambda \sim 630$ m. For smaller amplitudes we have of course bigger wavelengths. We take $h=200$ m and $\gamma+ 2\omega=0.1$ rad s$^{-1}$ (which is a typical vorticity value that can be observed in the upper layer in the ocean).  Rescaling back $X,$ $\theta$ and $\eta$ to their original physical values $(X\rightarrow X/\varepsilon, \theta\rightarrow R(X)-t)$ we re-write equation \eqref{KdV} in the form

\begin{equation} \label{KdV1}
bc^2 \left(2c+ (\gamma +2\omega)b\right)\eta_X
+
b c^2 c_X \eta
+ \frac{b^3}{3}\eta_{\theta\theta\theta}  + 
  (3c^2+3\gamma bc+\gamma^2 b^2)   \eta\eta_{\theta}
=0.
\end{equation}
The wave speed we determine from equation \eqref{c} taking the positive branch as the right-running waves:

\begin{equation}
c(X)=\frac{1}{2}\left(-(\gamma+2\omega) b(X) +\sqrt{(\gamma+2\omega)^2 b^2(X) +4gb(X) }\right).
\end{equation}

\begin{figure}[h]
\begin{center}
\includegraphics[totalheight=0.4\textheight]{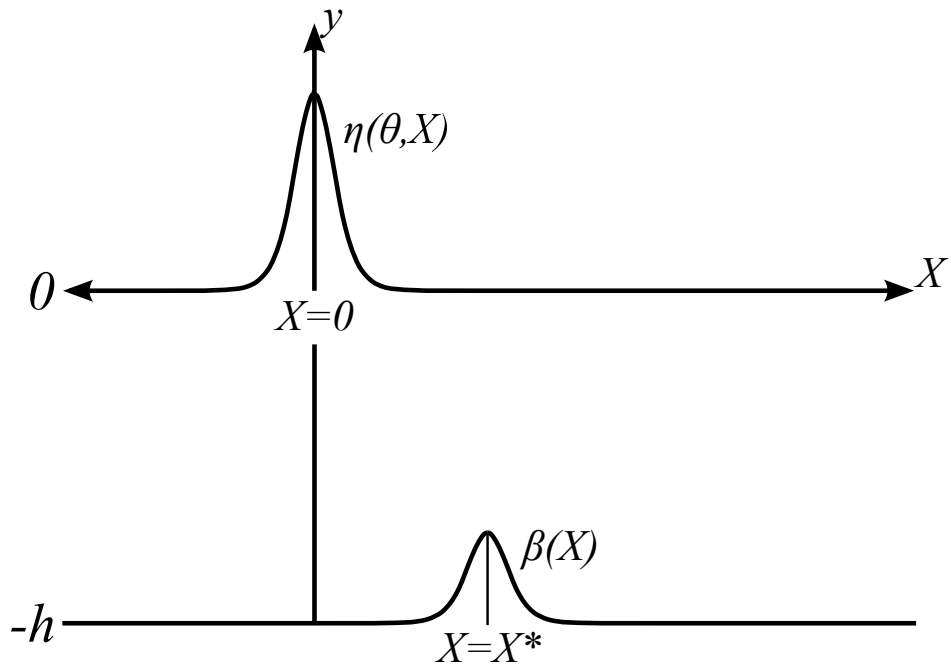}
\caption{Bottom obstacle as a ``hump''.}
\label{fig:figure_hump}
\end{center}
\end{figure}

\begin{figure}[h!]
\begin{center}
\fbox{\includegraphics[totalheight=0.3\textheight]{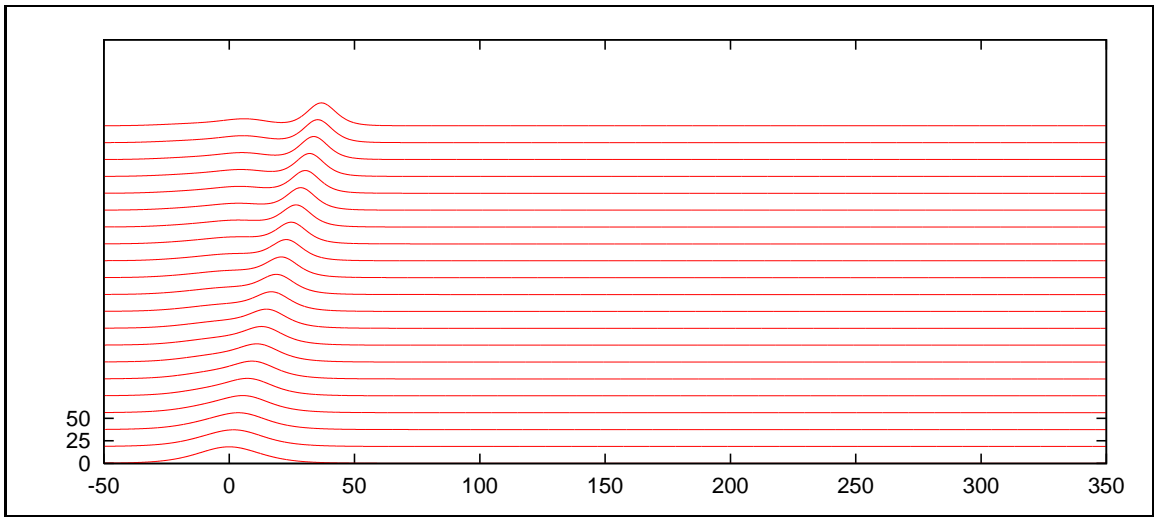}}
\caption{The parameter $Q=0.5$ with  a solitary wave moving above an obstacle. }
\label{fig:q=05}
\end{center}
\end{figure}

\begin{figure}[h!]
\begin{center}
\fbox{\includegraphics[totalheight=0.3\textheight]{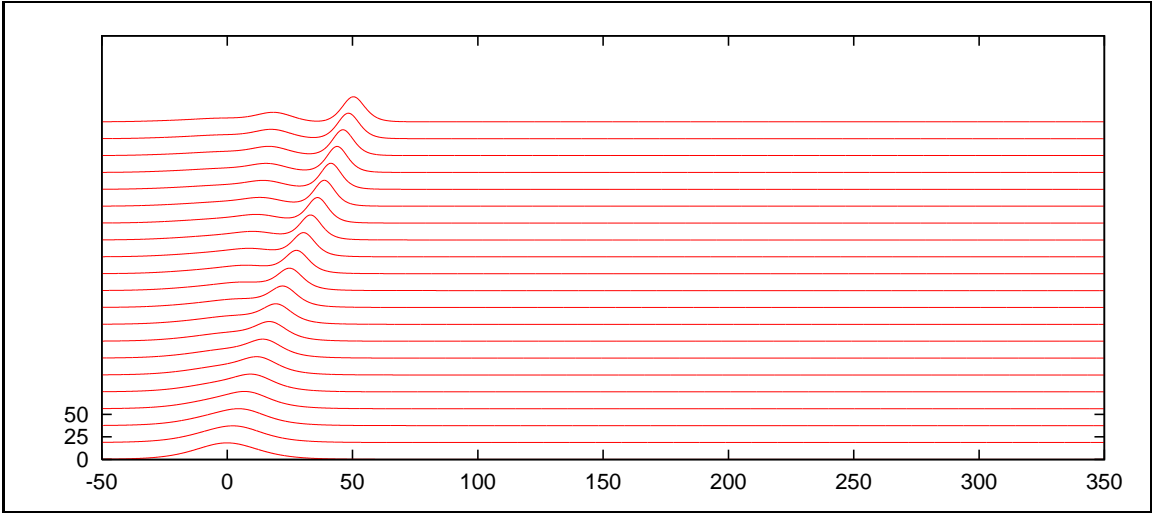}}
\caption{The parameter $Q=0.6$ with a second emerging soliton. A small amplitude (and velocity) soliton is visible behind the bigger one.}
\label{fig:q=06}
\end{center}
\end{figure}

\begin{figure}[h!]
\begin{center}
\fbox{\includegraphics[totalheight=0.3\textheight]{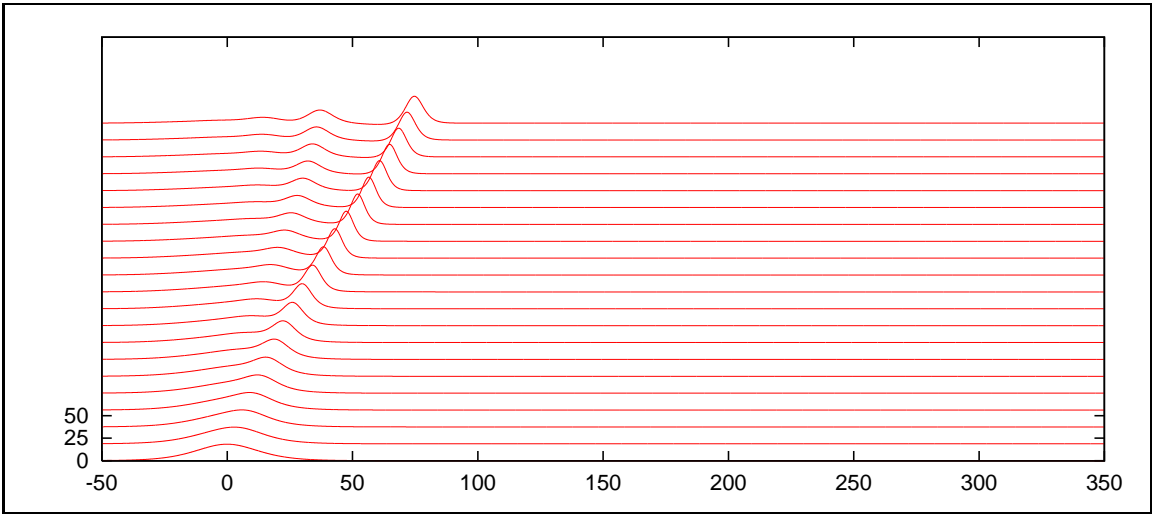}}
\caption{The parameter $Q=0.7$ with  2 emerging solitons.}
\label{fig:q=07}
\end{center}
\end{figure}

\begin{figure}[h!]
\begin{center}
\fbox{\includegraphics[totalheight=0.3\textheight]{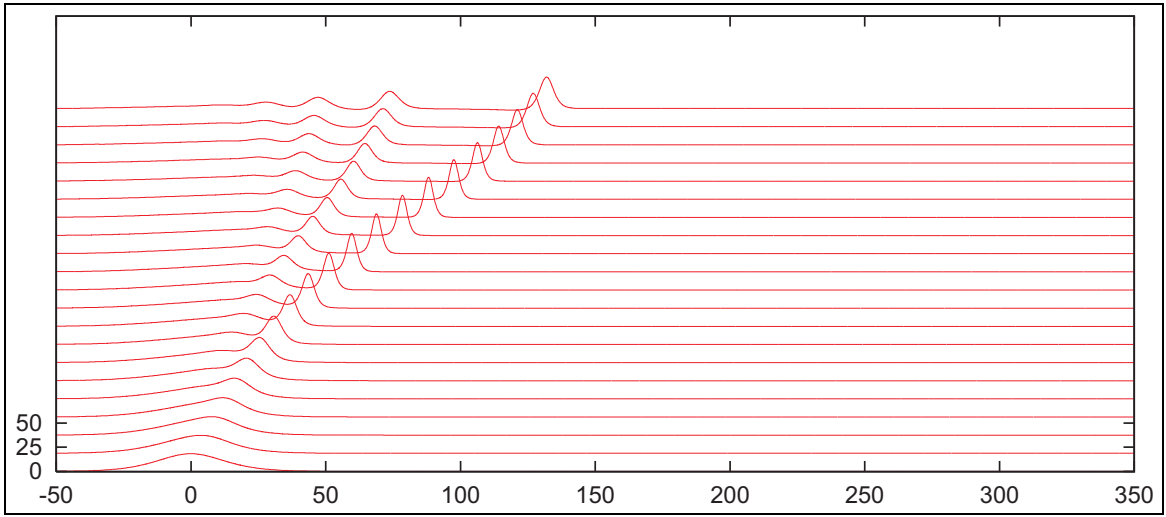}}
\caption{The parameter $Q=0.8$ with three emerging solitons. }
\label{fig:q=08}
\end{center}
\end{figure}

\begin{figure}[h!]
\begin{center}
\fbox{\includegraphics[totalheight=0.3\textheight]{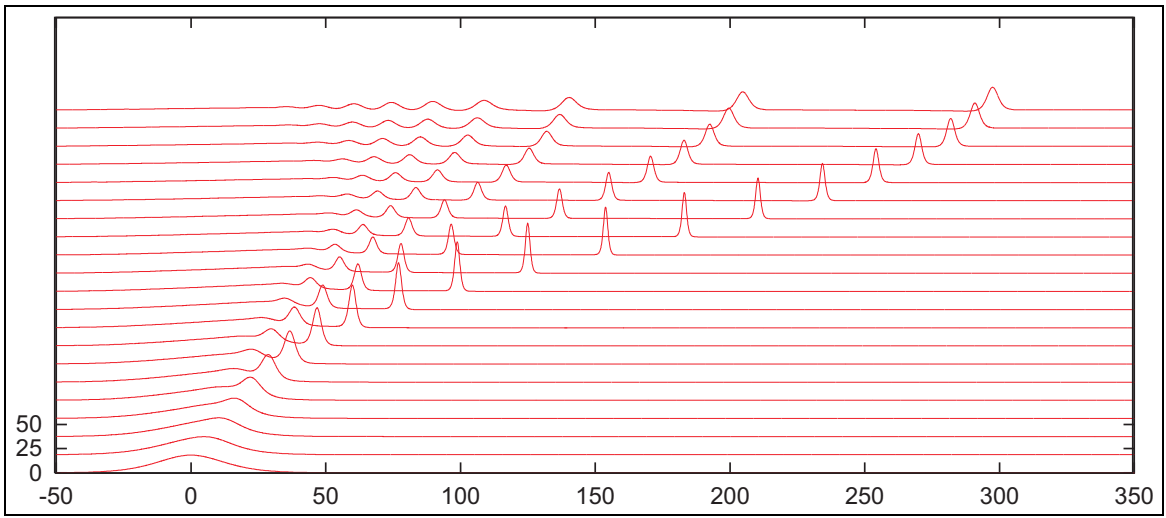}}
\caption{The parameter $Q=0.9$ with several emerging solitons. }
\label{fig:q=09}
\end{center}
\end{figure}

We assume that the wave approaches some sort of obstacle at $X=X^*$ 

$$b(X)=h\left[1-Q \exp \left(-\mu \left(\frac{X-X^*}{h}\right)^2\right)\right]$$ 
that is $$\beta(X)= Q \exp \left(-\mu \left(\frac{X-X^*}{h}\right)^2\right)$$

\noindent with $Q<1$ measuring the height of the obstacle and $\mu \ll 1$ measuring the width of the obstacle, see Fig. \ref{fig:figure_hump}. 
Far from the obstacle, at $X=-\infty$ the bottom has a depth $y=-h$ and the wave speed for the right running waves is
\begin{equation}
c_0=\frac{1}{2}\left(-(\gamma+2\omega) h +\sqrt{(\gamma+2\omega)^2 h^2 +4gh }\right).
\end{equation}
Equation \eqref{KdV1} in this limit becomes 

\begin{equation}\label{kdv1}
hc_0^2 \left(2c_0+ (\gamma +2\omega)h\right)\eta_X
+\frac{h^3}{3}\eta_{\theta\theta\theta}  + 
  (3c_0^2+3\gamma hc_0+\gamma^2 h^2)   \eta\eta_{\theta}
=0.
\end{equation}
This is clearly the KdV equation whose one-soliton solution is 
\begin{equation}\label{1sol} 
\eta(X,\theta)= \frac{A^2h^3}{3c_0^2+3h c_0\gamma + h^2 \gamma^2} \, \text{sech}^2 \left\{\frac{A}{2}\left(\theta-\frac{A^2h^2}{3c_0^2(2c_0+h(\gamma+2\omega))}X\right)\right\}
\end{equation}
where $A$ is a constant. For $A=0.12$ s$^{-1}$, $c_0=35$ m s$^{-1}$ the amplitude for $\eta$ given by the expression multiplying $\text{sech}^2$ is $18.6$ m. In our numerical studies we are using the one-soliton solution \eqref{1sol} as an initial condition for equation (\ref{kdv1}). The parameters for all cases presented in Figs. \ref{fig:q=05} to \ref{fig:q=09} are the same ($h=200$ m, $\gamma+ 2\omega=0.1$ rad s$^{-1}$, $\mu=0.01$, $A=0.12$ s$^{-1}$, $c_0=35$ m s$^{-1}$,  $X^*=1000$ m) only the parameter $Q$ that measures the height of the obstacle is different. The numerical results are the fully implicit finite-difference implementation of \eqref{kdv1} complemented by an inner iteration with respect to the nonlinear term (for more details see, for example, Todorov and Christov (2007)).
In all Figs. \ref{fig:q=05} to \ref{fig:q=09} the wave elevation $\eta$ is shown. The horizontal axis is the (slow) characteristic variable $\theta=R(X)-t $ measured in seconds. The snapshots are taken for different $X$ (recall that $X$ now coincides with $x$ numerically but it is also the time-like variable, see equation \eqref{t}). It is shown how the wave profile is changing with time as the initial KdV soliton travels though the region with a hump at the bottom. We observe emerging solitons and reflected rapidly decaying waves of radiation. Of course there is a change in the propagation speed when the incoming wave passes above the obstacle, due to the dependence of $c$ on $X$. The emerging solitons are smaller by amplitude and therefore by energy in comparison to the incoming wave. This is an example of how a reef prevents the propagation of waves with a big amplitude. The increase of the emerging solitons with $Q$ can be explained by means of the spectral theory of the KdV equation, as in the works by Johnson (1973a,b) or in the review article by Compelli et al (2018). It is observed that the dispersive radiation waves of small amplitude move to the left. This is because their phase velocity determined from the linearised equation $c^2 \left(2c+ (\gamma +2\omega)\right)\eta_X
+ \frac{b^2}{3}\eta_{\theta\theta\theta}=0 $ when $2c+ (\gamma +2\omega)h>0$ is $$ - \frac{b^2 k^2}{3c^2(2c+ (\gamma +2\omega)b)} k^2 <0$$ and becomes significant for the short waves where the wave number $k$ is not small. This effect is unphysical since the KdV model does not work as a water-wave model for short waves. The soliton's velocity, on the other hand, is positive as can be seen from equation \eqref{1sol}.

\section{Discussion}

\hspace{\parindent}The motion of the wave surface is determined by the variables $\eta(x,t)$ and the potential on the surface $\phi(x,t)$. The fluid potential $\varphi(x,y,t)$ in the fluid body can be recovered from $\phi(x,t)$ (Craig et al., 2005a) and then the velocity field in the fluid body from equation \eqref{stream_and_potential}. 
In many cases there are internal waves which form below the surface, between the top layer of warm water and the bottom layer of cold water with high salinity and higher density. There are by now numerous studies of a fluid system of two discrete fluid layers separated by an internal wave. We mention only the important works of Benjamin and Bridges (1997a,b), Craig and Groves (2000), Craig et al. (2005b). Wave-current interactions in a two-layer system have been investigated by Constantin and Ivanov (2015),  Constantin et al. (2016), Compelli and Ivanov (2015, 2017), Compelli (2016) and others. Therefore a system with both surface and internal waves over uneven bottom is of particular interest for future studies.
\\
%Solitary waves have been studied in Constantin et al. (2011).
%In Craig et al. (2006) the nonlinear interaction of solitary waves %are considered. Recent relevant studies of the KdV equation and other long-wave %models include
% Craig et al. (2005b), Craig and Groves (1994), Wahlen (2008).

\subsubsection*{Acknowledgements} 

\hspace{\parindent}AC, RI and CM are grateful to the Erwin Schr\"odinger International Institute for Mathematics and Physics (ESI), Vienna (Austria)  for the opportunity to work on the Research in Teams project {\it Hamiltonian approach to modelling geophysical waves and currents with impact on natural hazards} (2017) and to  participate in the programme {\it Mathematical Aspects of Physical Oceanography} (2018) where a significant part of this work has been accomplished. AC is supported by a Fiosraigh fellowship at Dublin Institute of Technology (Ireland). AC and CM are also funded by SFI grant 13/CDA/2117. MT acknowledges financial support from the Bulgarian Science Fund under grant DFNI I-02/9.  All authors are very grateful to the anonymous referees for their important suggestions.

%\begin{thebibliography}{10}
%\addcontentsline{toc}{section}{References}
%g\bibliographystyle{alpha}
%\expandafter\ifx\csname url\endcsname\relax
%  \def\url#1{\texttt{#1}}\fi
%\expandafter\ifx\csname urlprefix\endcsname\relax\def\urlprefix{URL }\fi
%\expandafter\ifx\csname href\endcsname\relax
%  \def\href#1#2{#2} \def\path#1{#1}\fi
  
\section*{References}

Benjamin, T.B. and Olver, P.J., 1982, Hamiltonian structure, symmetries and conservation laws for water waves. J. Fluid Mech. 125, 137--185, http://dx.doi.org/10.1017/S0022112082003292.

\vspace{2mm}
\noindent
Benjamin, T.B. and  Bridges, T.J., 1997a, Reappraisal of the Kelvin-Helmholtz problem. Part 1. Hamiltonian structure.  J. Fluid Mech. 333, 301--325,
http://dx.doi.org/10.1017/S0022112096004272.

\vspace{2mm}
\noindent
Benjamin, T.B. and  Bridges, T.J., 1997b, Reappraisal of the Kelvin-Helmholtz problem. Part 2. Interaction of the Kelvin-Helmholtz, superharmonic and Benjamin-Feir instabilities. J. Fluid Mech. 333, 327--373, http://dx.doi.org/10.1017/S002211209gg6004284.

\vspace{2mm}
\noindent
Boussinesq, J., 1872, Th{\'e}orie des ondes et des remous qui se propagent le long d'un canal rectangulaire horizontal, en communiquant au liquide contenu dans ce canal des vitesses sensiblement pareilles de la
surface au fond. J. Math. Pures Appl. 17, 55--108.

\vspace{2mm}
\noindent 
Compelli, A. and Ivanov, R., 2015, On the dynamics of internal waves interacting with the Equatorial Undercurrent. J. Nonlinear Math. Phys. 22, 531--539, http://dx.doi.org/10.1080/14029251.2015.1113052; arXiv:1510.04096

\vspace{2mm}
\noindent
Compelli, A. 2016, Hamiltonian approach to the modeling of internal geophysical waves with vorticity. Monatsh. Math. 179(4), 509--521, http://dx.doi.org/10.1007/s00605-014-0724-1.

\vspace{2mm}
\noindent 
Compelli, A. and Ivanov, R., 2017, The dynamics of flat surface internal geophysical waves with currents. J. Math. Fluid Mech.
19(2), 329--344, http://dx.doi.org/10.1007/s00021-016-0283-4; \linebreak arXiv:1611.06581

\vspace{2mm}
\noindent
Compelli, A., Ivanov, R. and Todorov, M., 2018, Hamiltonian models for the propagation of irrotational surface gravity waves over a variable bottom. Phil. Trans. R. Soc. A 376,  20170091, http://dx.doi.org/10.1098/rsta.2017.0091;
arXiv:1708.06791

\vspace{2mm}
\noindent 
Constantin, A. and Escher, J., 2004, Symmetry of steady periodic surface
water waves with vorticity. J. Fluid Mech. 498, 171--181, http://dx.doi.org/10.1017/S0022112003006773.

\vspace{2mm}
\noindent 
Constantin, A. and Strauss, W., 2004, Exact steady periodic water waves with vorticity. Comm. Pure Appl. Math. 57, 481--527, http://dx.doi.org/10.1002/cpa.3046.

\vspace{2mm}
\noindent 
Constantin, A., Sattinger, D. and Strauss, W., 2006, Variational formulations for steady water waves with vorticity. J. Fluid Mech. 548, 151--163,
http://dx.doi.org/10.1017/S0022112005007469.

\vspace{2mm}
\noindent
Constantin, A., Ivanov, R. and Prodanov, E., 2008, Nearly-Hamiltonian structure for water waves with constant vorticity. J. Math. Fluid Mech. 10, 224--237,
https://doi.org/10.1007/s00021-006-0230-x; 

\vspace{2mm}
\noindent
Constantin, A., 2011, Nonlinear water waves with applications to wave-current interactions and tsunamis. \textit{CBMS-NSF Regional Conference Series in Applied Mathematics} \textbf{81} Society for Industrial and Applied Mathematics, Philadelphia,https://doi.org/10.1137/1.9781611971873.

\vspace{2mm}
\noindent 
Constantin, A. and Escher, J., 2011, Analyticity of periodic traveling free surface water waves with vorticity. Ann. Math. 173, 559--568, 
https://doi.org/10.4007/annals.2011.173.1.12.

\vspace{2mm}
\noindent
Constantin, A., 2013, On equatorial wind waves. Differential and Integral Equations 26, 237--252, 

\vspace{2mm}
\noindent 
Constantin, A. and Ivanov, R., 2015, A Hamiltonian approach to wave-current interactions in two-layer fluids. Phys. Fluids 27, 08660, https://doi.org/10.1063/1.4929457.

\vspace{2mm}
\noindent 
Constantin, A. and Johnson, R.S., 2015, The dynamics of waves interacting with the Equatorial Undercurrent. Geophys. Astrophys. Fluid Dyn. 109(4), 311--358, https://doi.org/10.1080/03091929.\linebreak 2015.1066785.

\vspace{2mm}
\noindent
Constantin, A., Ivanov, R. and Martin, C.-I., 2016, Hamiltonian formulation for wave-current interactions in stratified rotational flows. Arch. Rational Mech. Anal. 221(3), 1417--1447, https://doi.org/ \linebreak 10.1007/s00205-016-0990-2.

\vspace{2mm}
\noindent
Constantin, A. and  Johnson, R.S., 2016a, An exact, steady, purely azimuthal equatorial flow with a free surface. J. Phys. Oceanogr. 46, no. 6, 1935--1945.

\vspace{2mm}
\noindent
Constantin, A. and Johnson, R.S.,  2016b, An exact, steady, purely azimuthal flow as a model for the Antarctic Circumpolar Current. 
J. Phys. Oceanogr., 46, no. 12, 3585--3594. 

\vspace{2mm}
\noindent
Constantin, A. and Johnson, R.S., 2017a,  Large gyres as a shallow-water asymptotic solution of Euler's equation in spherical coordinates. Proc. Roy. Soc.  A. 473, no. 2200, 20170063 (17 pp.), https://doi.org/10.1098/rspa.2017.0063.

\vspace{2mm}
\noindent
Constantin, A. and Johnson, R.S., 2017b, A nonlinear, three-dimensional model for ocean flows, motivated by some observations of the Pacific Equatorial Undercurrent and thermocline. Phys. Fluids 29(5), 056604, https://doi.org/10.1063/1.4984001.

\vspace{2mm}
\noindent
Constantin, A. and  Monismith, S.G., 2017, Gerstner waves in the presence of mean currents and rotation, J. Fluid Mech. 820, 511-528, https://doi.org/10.1017/jfm.2017.223.

\vspace{2mm}
\noindent
Craig,  W. and  Groves,  M., 1994,  Hamiltonian  long-wave  approximations  to  the  water-wave  problem. Wave Motion 19, 367--389, https://doi.org/10.1016/0165-2125(94)90003-5.

\vspace{2mm}
\noindent
Craig, W. and Groves, M., 2000, Normal forms for waves in fluid interfaces. Wave Motion 31, 21--41, https://doi.org/10.1016/S0165-2125(99)00022-0.

\vspace{2mm}
\noindent
Craig, W., Guyenne P., Nicholls D.P. and  Sulem C., 2005a, Hamiltonian long-wave expansions for water waves over a rough bottom. Proc. Roy. Soc. A  461, 839--873 https://doi.org/10.1098/rspa.2004.1367.

\vspace{2mm}
\noindent 
Craig, W.,  Guyenne, P. and Kalisch, H., 2005b, Hamiltonian long-wave expansions for free surfaces and interfaces. Comm. Pure Appl. Math. 58, 1587--1641,
https://doi.org/10.1002/cpa.20098.

\vspace{2mm}
\noindent
Craig, W.,  Guyenne, P., Hammack, J., Henderson, D. and Sulem, C., 2006, Solitary water wave interactions. Phys. Fluids 18, 057106,
https://doi.org/10.1063/1.2205916.

\vspace{2mm}
\noindent 
de Bouard, A.,  Craig, W.,  D{\'i}az-Espinosa, O., Guyenne, P. and Sulem, C., 2008, Long wave expansions for water waves over random topography. Nonlinearity 21(9), 2143--2178, https://doi.org/10.1088/0951-7715/21/9/014.

\vspace{2mm}
\noindent
Dejak, S.I. and Sigal, I.M., 2006, Long-time dynamics of KdV solitary waves over a variable bottom. Comm. Pure Appl. Math. 59(6), 869--905,
https://doi.org/10.1002/cpa.20120.

\vspace{2mm}
\noindent
Fedorov A.V. and Brown, J.N., 2009, Equatorial waves.
Encyclopedia of Ocean Sciences, edited by J. Steele, pp. 3679--3695, Academic Press: New York.

\vspace{2mm}
\noindent 
Escher E., Henry, D., Kolev, B. and  Lyons, T., 2016, Two-component equations modelling water waves with constant vorticity. Ann. Mat. Pura Appl. 195(1), 249--271, https://doi.org/10.1007/s10231-014-0461-z.

\vspace{2mm}
\noindent
Henry, D., 2013, An exact solution for equatorial geophysical water waves with an underlying current. Europ. J. Mech. B/Fluids 38, 18--21,
https://doi.org/10.1016/j.euromechflu.2012.10.001.

\vspace{2mm}
\noindent
Ionescu-Kruse, D. and  Martin, C.-I., 2017, Periodic equatorial water flows from a Hamiltonian perspective. J. Diff. Equations 262, no. 8, 4451--4474, https://doi.org/10.1016/j.jde.2017.01.001.

\vspace{2mm}
\noindent
Johnson, R.S., 1997, \textit{Mathematical Theory of Water Waves}. Cambridge Texts in Applied Mathematics.

\vspace{2mm}
\noindent 
Johnson, R.S., 1973a, On an asymptotic solution of the Korteweg-de Vries equation with slowly varying coefficients. J. Fluid Mech. 60,  813--824
https://doi.org/10.1017/S0022112073000492.

\vspace{2mm}
\noindent
Johnson, R.S., 1973b, On the development of a solitary wave moving over an uneven bottom. Proc. Camb. Phil. Soc. 73, 183--203,
https://doi.org/10.1017/S0305004100047605.

\vspace{2mm}
\noindent
Johnson, R.S., 2018, Application of the ideas and techniques of classical fluid mechanics to some problems in physical oceanography. Phil. Trans. Roy. Soc. A  376,  20170092, https://doi.org/10.1098/\linebreak rsta.2017.0092. 

\vspace{2mm}
\noindent
Jonsson, I.G., 1990,  Wave-current interactions. In: \textit{The sea} 9(3A), Wiley New York 65--120.

\vspace{2mm}
\noindent  
Kaup, D.J., 1975, A Higher-order water-wave equation and the method for solving it. Progr. Theor. Phys. 54(2),396--408, https://doi.org/10.1143/PTP.54.396.

\vspace{2mm}
\noindent  
Korteweg, D.J. and Vries, G., 1895,  On the change of form of long waves advancing in a rectangular channel and on a new type of long stationary wave. Philos. Mag. 39, 422--443, https://doi.org/10.1080/ \linebreak 14786449508620739.

\vspace{2mm}
\noindent
Martin, C.-I., 2018, On periodic geophysical water flows with discontinuous vorticity in the equatorial $f$-plane approximation. Philos. Trans. Roy. Soc. A 376, no. 2111, 20170096 (23 pp), https://doi.org/10.1098/rsta.2017.0096.

\vspace{2mm}
\noindent 
Milder, D., 1977, A note regarding ``On Hamilton's principle for water waves''. J. Fluid Mech. 83, 159--161, https://doi.org/10.1017/S0022112077001116.

\vspace{2mm}
\noindent
Miles, J.W., 1977, On Hamilton's principle for water waves. J. Fluid Mech. 83(1), 153--158, \linebreak https://doi.org/10.1017/S0022112077001104.

\vspace{2mm}
\noindent
Nachbin, A., 2003, A Terrain-Following Boussinesq System. SIAM J. Appl. Math. 63(3), 905--922, https://doi.org/10.1137/S0036139901397583.

\vspace{2mm}
\noindent
Nachbin, A. and Ribeiro-Jr., R., 2018, Capturing the flow beneath
water waves. Phil. Trans. Roy. Soc. A 376, 20170098, https://doi.org/10.1098/rsta.2017.0098.
 
\vspace{2mm}
\noindent 
Peregrine D.H., 1976, Interaction of Water Waves and Currents. Advances in Applied Mechanics 16, 9--117, https://doi.org/10.1016/S0065-2156(08)70087-5. 

\vspace{2mm}
\noindent 
Teles da Silva, A.F. and Peregrine, D.H., 1988, Steep, steady surface waves on water of finite depth with constant vorticity. J. Fluid Mech. 195, 281--302, http://dx.doi.org/10.1017/S0022112088002423.

\vspace{2mm}
\noindent
Thomas, G.P., 1981, Wave-current interactions: an experimental and numerical study. Part 1. Linear waves, J. Fluid Mech. 110, 457--474, https://doi.org/10.1017/S0022112081000839. 

\vspace{2mm}
\noindent
Thomas, G.P., 1990,  Wave-current interactions: an experimental and numerical study. Part 2. Nonlinear waves. J. Fluid Mech. 216, 505--536, https://doi.org/10.1017/S0022112090000519. 

\vspace{2mm}
\noindent
Thomas, G.P. and Klopman, G., 1997, Wave-current interactions in the near-shore region, in Gravity Waves in Water of Finite Depth, J. N. Hunt, Ed., Advances in Fluid Mechanics, pp. 255--319, Computational Mechanics Publications. 

\vspace{15mm}
\noindent 
Wahl{\'e}n, E., 2007, A Hamiltonian formulation of water waves with constant vorticity. Lett. Math. Phys. 79, 303--315, http://dx.doi.org/10.1007/s11005-007-0143-5.

\vspace{2mm}
\noindent
Todorov, M.D. and Christov, C.I., 2007 Conservative numerical scheme in complex arithmetic for Coupled  Nonlinear Schrodinger Equations. Discrete Contin. Dyn. Syst. {\bf Suppl. 2007}, 982--992.

\vspace{2mm}
\noindent
Zakharov, V.E., 1968, Stability of periodic waves of finite amplitude on the surface of a deep fluid. Zh. Prikl. Mekh. Tekh. Fiz. 9, 86--94 (in Russian); J. Appl. Mech. Tech. Phys. 9, 190--194 (English translation)
http://dx.doi.org/10.1007/BF00913182.

%\end{thebibliography}

g

\end{document}